\documentstyle[12pt,aasms4]{article}

\slugcomment{Submitted to ApJ: 26 January 1999; Accepted: 28 April 1999}

\begin{document}

\title{Gas Dynamics in the Luminous Merger NGC 6240}

\author{L.J.Tacconi, R.Genzel, M.Tecza and
J.F.Gallimore\altaffilmark{1}}
\affil{Max-Planck Institut f\"ur extraterrestrische Physik (MPE), D-85748
Garching, Germany}

\author{D.Downes}
\affil{Institut de Radio Astronomie Millim\'etrique (IRAM), 38406 
St. Martin d'H\`eres, France}

\and

\author{N.Z. Scoville}
\affil{California Institute of Technology, 105-24, Pasadena, CA 91125}

\altaffiltext{1}{present address: National Radio Astronomy Observatory,
Charlottesville, VA  22903}

\newcommand{\degr}{^{\circ}\ }
\newcommand{\degrn}{^{\circ}}
\newcommand{\etal}{{\em et al.\/~}}
\newcommand{\amin}{^{\prime}}
\newcommand{\rar}{\rightarrow}
\newcommand{\asec}{^{\prime\prime}}
\newcommand{\kms}{\rm km\  s^{-1}}

\begin{abstract}
We report $0\farcs5\times 0\farcs9$ resolution,
interferometric observations of the 1.3 mm CO J=2$\rightarrow$1 line in the
infrared luminous galactic merger NGC~6240 \footnote{Based on observations 
carried out with the IRAM Plateau de Bure Interferometer.  IRAM is supported 
by INSU/CNRS (France), MPG (Germany) and IGN (Spain).}. About half of the CO 
flux is concentrated in a rotating but highly turbulent, thick disk structure
centered between the two radio and near-infrared nuclei. A number of gas
features connect this $\sim$500 pc diameter central disk to larger scales.
Throughout this region the molecular gas has local velocity widths which
exceed 300 $\kms$ FWHM and even reach FWZP line widths of 1000 $\kms$ in a
number of directions. The mass of the central gas concentration constitutes
a significant fraction of the dynamical mass, M$_{gas}$(R$\le$470 pc) $\sim$
2-4$\times$10$^9$ M$_{\odot}$ $\sim$ 0.3-0.7 M$_{dyn}$. We conclude that
NGC~6240 is in an earlier merging stage than the prototypical ultraluminous
galaxy, Arp~220. The interstellar gas in NGC~6240 is in the process of
settling between the two progenitor stellar nuclei, is dissipating rapidly
and will likely form a central thin disk. In the next merger stage, NGC~6240
may well experience a major starburst like that observed in Arp~220.
\end{abstract}

\section{Introduction}

In the last two decades it has become clear that collisions and mergers of
galaxies are important processes that trigger spectacular starburst and
nuclear activity, and that play a critical role in the evolution of galaxies
(e.g. Sanders and Mirabel 1996, Genzel, Lutz and Tacconi 1998). Hierarchical
merging of smaller sub-units into bigger ones was probably a key process in
the formation of galaxies. It is, therefore, of obvious importance to study
and understand in detail some prototypical examples of nearby merger systems.

The galaxy NGC~6240 (D=97~Mpc for H$_{o}$ = 75 $\kms$ Mpc$^{-1}$, 
1$^{\prime \prime } $ = 470 pc) belongs to the class of [ultra]-luminous 
infrared galaxies
([U]LIRGs) whose radiation output largely emerges in the infrared band (L$
_{bol}$(NGC~6240) $\sim $ L$_{IR}$ = L$_{8-1000\mu m}$ = 6$\times $10$^{11}$
L$_{\odot }$, Sanders et al. 1988, Thronson et al. 1990, Sanders and Mirabel
1996). The main characteristics of NGC~6240 are:

\begin{itemize}
\item  {Its optical/near-infrared/radio morphology is dominated by two
compact sources (hereafter referred to as `nuclei') at a separation of $\sim
1\farcs6$(750 pc), surrounded by a highly disturbed stellar disk
structure along P.A. = 25$^{\circ}\ $ E of N (Fried and Schulz 1983,
Thronson et al. 1990, Colbert et al. 1994). The large de-reddened K-band
luminosities \footnote{$L(K) \simeq (\nu L_{\nu})_{2.2 \mu m} \times (\Delta
\lambda / \lambda)_{K-band}$, where the fractional width of the K-band is $
(\Delta \lambda / \lambda)_{K-band} \sim 0.23$.} of these two main emission
peaks (L(K)$\sim 2\times10^9 L_{\odot}$ for the south and $\sim 1\times10^9
L_{\odot}$ for the north nucleus) show that they must be massive: $\ge
1\times 10^9 M_{\odot}$ and $\ge 5\times10^8 M_{\odot}$ for the south and
north nuclei, respectively. They are more than just bright star forming
regions and are very plausibly individual galaxy nuclei. NGC~6240, thus, is
very probably a system of colliding/merging galaxies that we are viewing
prior to the final merging of the two progenitor nuclei into one.}

\item  {The infrared to radio continuum spectral energy distribution and the
mid-/far-infrared line emission indicates that a large fraction of the
bolometric luminosity of NGC~6240 comes from a recent but aging starburst
(Colbert et al. 1994, Lutz et al. 1996, Genzel et al. 1998). In addition,
5-10 keV X-ray continuum emission, 6.7 keV Fe K$\alpha$ emission ($
L_x/L_{IR} \sim 10^{-2}$, Kii et al. 1997, Nakagawa et al. 1999) and fairly
strong 26$\mu$m [OIV] mid-IR line emission (Lutz et al. 1996, Genzel et al.
1998, Egami et al. 1999) all strongly suggest that an AGN is present and
probably significantly contributes to L$_{bol}$.}

\item  {Ionized gas with velocity widths of up to 1000 $\kms$ is observed in
optical emission lines, such as H$\alpha$ (Heckman, Armus and Miley 1987,
1990; Armus, Heckman and Miley 1990, Bland-Hawthorn et al. 1991). Heckman,
Armus and Miley (1987) and Armus, Heckman and Miley (1990) interpret the
dynamics and hour-glass shaped morphology of the H$\alpha$ emission (along
P.A. = 110$^{\circ}\ $, perpendicular to the large scale disk) in terms of a
powerful superwind triggered by the nuclear activity and supernova
explosions. The ratio of kinetic energy input by the wind to bolometric
luminosity is about 1.5\% (Heckman et al. 1990).}

\item  {NGC~6240 has the most powerful infrared H$_{2}$ line emission so far
found in a galaxy (Joseph et al. 1984; L(H$_{2}$)$\sim 2\times 10^{9}$ L$
_{\odot }$: Egami et al. 1999). The 2$\mu $m H$_{2}$ v=1$\rightarrow $0 S(1)
emission is centered between the two radio/infrared nuclei (Herbst et al.
1990; van der Werf et al. 1993; Tecza et al. 1999) and probably is excited
in shocks due to the galaxy collision (van der Werf et al. 1993). Intense
millimeter CO rotational line emission indicates that $\sim 10^{10}$ M$
_{\odot }$ of dense molecular interstellar gas is concentrated in the
central kiloparsec (Wang, Scoville and Sanders 1991, Solomon, Downes and
Radford 1997, Bryant and Scoville 1999). The large amount of molecular gas 
near the nuclei implies
large equivalent column densities of dust (N(H$_{2}$)$\geq 2\times 10^{23}$
cm$^{-2}$ or A$_{V}$$\geq $200). It also suggests that gas constitutes a
significant fraction of the dynamical mass.}

\item  {The late type stars in the region of the double nucleus have a very
large velocity dispersion, $\sigma =350\pm 20\ \kms$ in $1\farcs5$ to 
$2\farcs2$ diameter apertures (Lester and Gaffney 1994, Doyon et al. 1994). 
This dispersion is much larger than that observed in gas-rich, disk galaxies, 
and is at the upper end of what is observed for massive elliptical galaxies
(Doyon et al. 1994, Whitmore, McElroy and Tonry 1985, Bender et al. 1994).
The ratio of dynamical mass to dereddened total K-band stellar luminosity for
the N6240 system is
M(dyn)/L(K) = 4.5 for a K-band screen extinction of A(K)=0.5--0.7 (Genzel et
al. 1998). For a single stellar population this ratio is fit by an age 
$\sim$1 Gyr, assuming that the dynamical mass is due to the stars dominating 
the near-infrared light, that star formation is constant with time, and that 
the initial mass function is Scalo (1986) down to a lower mass cutoff of 
0.2 M$ _{\odot }$. This age is inconsistent with the equivalent width of the 
CO bandhead absorption. The latter implies that the K-band light is dominated
by massive red supergiants of age 10--30 million years (Lester et al. 1988,
Tecza et al. 1999). The discrepancy can only be resolved if the main carrier
of the dynamical mass does not contribute to the near-infrared light.}
\end{itemize}

To investigate the spatial distribution and dynamics of the neutral gas
component in this prototypical merger system we have carried out high
resolution imaging spectroscopy of the CO J = 2$\rightarrow$1 line at 1.3 mm
with the IRAM mm-interferometer on the Plateau de Bure. Our new data shed
light on the merger dynamics and evolution.

\section{Observations and Data Reduction}

We have mapped the $^{12}$CO J=2$\rightarrow $1 line in NGC~6240 with the
IRAM millimeter interferometer on the Plateau de Bure, France (Guilloteau et
al. 1992). The data were obtained between January and February 1998. The
array consisted of 5 15- meter antennas positioned in 3 different
configurations providing 30 baselines ranging from 32 to 408 meters. Since
the CO line emission is $\sim $1000 $\kms$ wide, we observed with two
different local oscillator setups centered $\pm $250 $\kms$ from the assumed
systemic velocity of 7339 $\kms$, providing an equivalent bandwidth of 970 MHz
(1300 $\kms$). We observed NGC~6240 for $\sim $6 hours in each configuration
and LO setting for a total of about 36 hours of integration time (18 hours
integration per velocity channel). Each antenna was equipped with
dual-frequency SIS receivers enabling us to observe the HCN J=1$\rightarrow$0 
line at 86.5 GHz simultaneously with the CO J=2$\rightarrow $1 line. The
SIS mixers have receiver temperatures of 60--80 K, and are used in
single-sideband mode at 3 mm and double-sideband mode at 1 mm.  The system
temperatures, corrected to outside the atmosphere, are 120 K and 400 K 
respectively. The phases and amplitudes were
calibrated by observing 1655+077 every half hour, and the bandpass was
calibrated by observations of 3C273. The absolute flux scale is based on a
flux of 14~Jy for 3C273 at 225 GHz. Fluxes of strong sources are determined
through careful monitoring with both the interferometer and the IRAM
30-meter telescope on Pico Veleta, Spain. The data are of excellent quality;
based on the monitoring measurements, the accuracy of the flux scale is
expected to be better than 30\%. The phase noise on the longest baselines
was 20$\degrn$--30$\degrn$ rms at 1 mm. Spectral resolution of 2.5 MHz 
($\sim $3.3
$\kms$ at 1 mm and 6.5 $\kms$ at 3 mm) was provided by 6 autocorrelator
spectrometers for each of the two LO frequency settings, covering the total
970 MHz bandwidth. The data were calibrated using the CLIC software package
written at IRAM. We then made and CLEANed uniformly weighted channel maps
using software from the GILDAS package. To increase the sensitivity we
smoothed the data spectrally to a resolution of 20 $\kms$ before mapping. The
CLEANed CO maps were reconvolved with a $0\farcs9 \times 0\farcs5$
FWHM beam (P.A.=26$^{\circ }\ $), and the HCN 1$\rightarrow $0 maps were 
reconvolved with a $2\farcs5 \times 1\farcs4$ FWHM beam (P.A.=29$^{\circ }\ $).
The rms noise (per channel) after CLEANing was $\sim $4 mJy beam$^{-1}$ for 
the CO 2$\rightarrow$1 data and 0.8 mJy beam$^{-1}$ for the HCN 
1$\rightarrow$0 maps.

We also observed NGC~6240 using the six element Multi-Element Radio-Linked
Interferometer (MERLIN) based at Jodrell Bank (Wilkinson 1992).
The observing frequency was 5~GHz with a bandwidth of 15~MHz. We observed two
tracks: one horizon-limited track on 7 Dec 1997 and a shorter track on
9 Dec 1997, for a total on-source integration time of roughly 11
hours. Phase calibration solutions were deteremined by interleaved
scans of the phase calibrator B1648+015 with a duty cycle of six
minutes source to one minute calibrator. The flux scale was set by the
daily scan of 3C~286 bootstrapped to the phase calibrator. Initial
calibration and editing employed the Jodrell Bank {\sc dprog}
software package.
We used standard fast Fourier transform imaging software in the AIPS
software package provided by NRAO \footnote{The National Radio
Astronomy Observatory is operated by Associated Universities, Inc.,
under contract with the National Science Foundation.} Two iterations
of self-calibration improved the phase calibration solutions for the
source. The resolution of the naturally-weighted synthetic map is $94
\times 50$~milliarcseconds into P.A. 21$\degrn$, and the image sensitivity
is 0.1~mJy beam$^{-1}$. 

\section{Results}

The basic results of our CO 2$\rightarrow $1 interferometric observations of
NGC~6240 are shown in Figures 1 through 6. Figure 1 shows selected spectra
and a logarithmic image of the integrated CO line emission. Figure 2 shows 
uniformly
weighted maps in $\Delta $v = 40 $\kms$ resolution velocity channels. Figure 3
displays a 228 GHz continuum map made from the upper sideband, and a
superposition of red- (v$\geq 350\ \kms$) and blue-shifted (v$\leq -350
\ \kms$) line wing emission. Figure 4 gives the second moment map of the CO
emission (intensity weighted measure of dispersion around the mean), and
position-velocity diagrams of data and an appropriate disk model. Figure 5
displays selected spectra in the central concentration along with those of
the model disk. Figure 6 shows an overlay of integrated and
red-/blue-shifted CO maps on top of the 8 GHz radio continuum map of Colbert
et al. (1994) and the 5 GHz MERLIN map. Table 1 summarizes the measured 
fluxes and source parameters.  We next summarize our basic findings.

\subsection{Spatial distribution of the gas}

Most of the CO 2$\rightarrow $1 emission in NGC~6240 comes from the central 3
$^{\prime \prime }$ radius region ($\sim $1.5 kpc). We detect a total of
1220 Jy $\kms$ of CO flux from NGC~6240. It is difficult to say how much of
the single dish flux we detect, since published single dish spectra span
insufficient bandwidth to include enough continuum for proper subtraction.
For instance, the spectra of Combes et al. (1991) and Casoli et al. (1992)
span only 660 $\kms$ and they quote an integrated line strength of $\sim $
800 Jy $\kms$. For comparison, we measure $\sim $1000 Jy $\kms$ over the
same bandpass. The discrepancy
probably owes to the baseline subtraction in the single dish spectra. Since
the line covers the entire single dish band, baseline subtraction would
remove both continuum and an indeterminate amount of line flux. Solomon
et al.~(1997) measure a CO 1$\rar$0 integrated intensity of 69 K $\kms$
in a 22$\asec$ aperture in NGC 6240.  This converts to a 2$\rar$1 flux
of 1000 Jy $\kms$, assuming that $^{12}$CO 2$\rar$1/1$\rar$0 
ratio is 0.8 (Casoli et al.~1992).  It is thus
likely that we are recovering most of the CO line emission in our
interferometric map.

Of the total source line flux, 528 Jy $\kms$, or 43\% is concentrated in a $
\sim $1$^{\prime \prime }$ diameter thick disk-like structure located in
between the two radio/infrared nuclei (Figure 6). Such a distribution
was strongly suggested in the lower resolution CO map of Bryant and
Scoville (1999).  The aspect ratio of the
integrated CO distribution in Figure 1 is 1:2 (EW:NS), slightly larger than
the aspect ratio expected from the inclination of the large scale disk (1:3
for i$\sim 70\degrn -75^{\circ }$, Fried and Schulz 1983). The two nuclei are
located approximately at either end of this central gas disk. This is best
seen in Figure 6 (middle) which compares the 5 GHz MERLIN map 
with the integrated CO emission shown on a linear intensity scale.
The centroid of the integrated CO emission is somewhat closer to ($0\farcs6$
NNE of) the brighter, southern nucleus. 

Only the two main nuclei are detected by the MERLIN observations (Figure
6b); the extended radio structure in the VLA maps of Colbert et
al. (1994; Figure 6a) is mostly resolved below the sensitivity of the MERLIN
images. The extended radio structure surrounding the southern nucleus
matches well the sidelobe pattern of the aperture synthesis beam and
is probably an artifact of the imaging and calibration process.
There is no
compact radio (or infrared) continuum source at or near the position of the
CO peak, neither on the $0\farcs15$ resolution 8 GHz map of
Colbert et al., nor on the $0\farcs08 \times 0\farcs035$
5 GHz MERLIN map.
This and the fact that the total radio flux density from the two nuclei falls
right on the radio-far-infrared correlation for star forming galaxies if
one takes the total far-infrared flux from the IRAS data
(Colbert et al. 1994) strongly suggests {\it that there is no strong
luminosity source hidden at the center of the CO emission}. However, there
is a bridge of extended 8 GHz emission connecting the southern and northern
nuclei and approximately following the general outline of the CO emission.

The orientation of the long axis of the CO disk (P.A.$\sim $10$\degrn$
--20$\degrn$, Figure 1) is about the same as that of the large scale
galactic disk of NGC~6240 but somewhat tipped away from the kinematic major
axis as derived in this paper (see section 3.2).
There are several distinct gas features in the integrated and channel maps
surrounding this central structure and extending to spatial scales of 3$
^{\prime\prime}$--4$^{\prime\prime}$. These features can be best identified
in the channel maps (see below). A second, smaller CO peak (220 Jy $\kms$ of
the total flux of 1220 Jy $\kms$ is located 4$^{\prime\prime}$--5$
^{\prime\prime}$ NNE of the CO maximum. The second peak (the `northern
clump') was first seen in the 7$^{\prime\prime}$ CO 1$\rightarrow$0
interferometry map of Wang, Scoville and Sanders (1991). It does not
correlate with any feature in the radio continuum, infrared or optical maps.

The HCN 1$\rightarrow$0 emission is confined to the central $\sim$3$
^{\prime\prime}$ radius region of the CO peak. We detect a total of 12 Jy
$\kms$ from this region. 
The profiles and widths of the HCN lines
are very similar to those seen in CO when the two data sets are smoothed to
the $2\farcs5 \times 1\farcs4$ resolution of the HCN
map. This is an indication that the emission from both species arises from
the same dense gas throughout the region of the CO peak. Since the CO data
are of much higher spatial resolution the rest of the paper is devoted to
the CO results alone.

\subsection{Kinematics of the gas}

The neutral gas in NGC~6240 is highly disturbed, with unusually large
velocity widths almost everywhere in the central 3$^{\prime\prime}$. Typical
line widths ($0\farcs5 \times 0\farcs9$ beam) are FWHM 300--400 $\kms$ and 
FWZP 700--1000 $\kms$. In the region $\sim 0\farcs5 - 1\farcs5$ north of the 
peak even the $\pm$540 $\kms$ \footnote{
All velocities in the text are relative to v$_{hel}$ = 7340 $\kms$} range
covered by our two LO settings was not sufficient to include all of the line
emission (Figure 1). In this region there are red wings extending to at
least $+600$ $\kms$. A study of the spectra in Figure 1 and the channel maps
in Figure 2 reveals that there are several distinct kinematic components.

In the central disk the centroids/peaks of the highly blueshifted emission ($
-$250 to $-$480 $\kms$) and of the highly redshifted emission (300--550 $\kms$)
are separated by $0\farcs75$ along P.A.=40$^{\circ }\ $ (Figure 3 and Figure 
6, bottom). The average position of the blue- and red-shifted
emission centroids is coincident with the peak of the 2$\mu $m vibrationally
excited H$_{2}$ emission (Tecza et al. 1999) and is $\sim 0\farcs15$ S of the 
CO integrated flux peak. The most likely interpretation
of this large velocity gradient is rotation of the central disk with a line
of nodes along a P.A. of $30\degrn -45^{\circ }\ $. However, the line 
profiles in
the central structure all are singly peaked, rather than the double-horned
profiles characteristic of a rotating disk. To wipe out the classical
double-horn there are probably two effects at work: (1) the gas motions
could have a very large random component and (2), the rotation curve of the
disk could be steeply rising from the center outwards. These effects could
combine most effectively if most of the disk emission comes from the lower
velocity central part of the disk. In addition, radiative transport effects
in an optically thick disk may further decrease the contrast between the
optically thicker, high velocity edges and the thinner lower velocity parts
of the emission profile. We investigate possible models in more detail
below. It is very unlikely that the velocity gradient is caused by a bipolar
outflow. The molecular gas mass in the central disk significantly exceeds 10$
^{9}$ M$_{\odot }$ (see section 4). The kinetic energy of the outflow thus
would be $\geq $10$^{57.5}$ ergs. The dynamical lifetime of the present
central CO concentration would then be $\leq $10$^{6}$ years and the
mechanical energy deposition rate would be $\geq 10^{10.5}$ L$_{\odot }$ $
\geq $ 0.05 L$_{bol}$. This deposition rate exceeds the mechanical
luminosity of the superwind by a factor of 3 or more. These energy and time
scale considerations make an outflow scenario very improbable. In addition,
it is then unclear where the origin of the outflow is, since there is no
radio source at the center of the CO peak. In this scenario it is not
obvious why the two radio nuclei are located at either end of the disk.

There are a number of separate gas features connecting the central 500 pc to
the larger scale disk (indicated as letters A through E on Figure 1). These
features are also present in the 2$\mu$m H$_2$ maps of Herbst et al. (1990),
van der Werf et al. (1993) and Tecza et al. (1999). The two most prominent
of these (A, B) emanate from the southern edge of the disk and arch to the
SW and SE on either side of the southern nucleus (Figures 1 and 2). Both
features can be easily recognized on the integrated line maps as well as in
the channel maps (Figures 1 and 2), largely because they have very broad
lines (Figure 1). As in the central region the broad lines all along the
lengths of these features are in fact their most prominent characteristic.
Systematic velocity gradients along the features are comparatively small.
The gas in feature C leads from the center to the SE and is mainly
blueshifted ($-$250 to 0 $\kms$), with a possible connection to a lower
velocity long filament stretching toward the NE. Features D and E consist of
redshifted gas (100 to 500 $\kms$) with very wide lines. They connect the
northern edge of the central concentration to the NNE and NNW, with a
possible bridge to the northern clump at $\sim 200\ \kms$. Taken together
features A+B and D+E appear to form a `butterfly' pattern emanating from the
nuclear concentration to the north and south, straddling around the two
infrared/radio nuclei.

NGC~6240 is highly unusual because of these large chaotic velocities. None
of the several other ULIRGs/mergers studied so far with high resolution mm-
interferometry have these characteristics (e.g. Yun and Scoville 1995,
Bryant and Scoville 1997, Scoville et al. 1997, Downes and Solomon 1998,
Sakamoto et al. 1999). It is difficult to interpret the kinematics and gas
distribution as a simple, orbital motion or uniform flow. The most likely
interpretation of the enormous line widths is that they largely represent
line of sight projections in a complex, highly chaotic gas flow. This system
has many streamers with a wide range of angular momenta and orbital
parameters. In addition one would expect that the large velocity widths
cause shocks and dissipation. Dissipation of the cloud kinetic energy
through shocks naturally explains the luminous infrared molecular hydrogen
emission; recall that NGC 6240 is the most luminous H$_{2}$ line source
currently known (Rigopoulou et al. 1999, section 5.1).  The fact that
these motions still exist in the central regions of NGC~6240 argues for
this system being in a relatively early phase of merging before the
chaotic motions have been dissipated.

\subsection{Rotating disk model of the central disk}

In order to quantify the dynamical properties of the central CO disk we fit
the data by an axisymmetric turbulent, rotating disk (see
Tacconi et al. 1994). Recognizing that the gas distribution and kinematics
are, in detail, more complicated than a simple axisymmetric model, the main
disk parameters were tuned interactively to fit the principal
characteristics of the data and to represent the axisymmetrically averaged
gas properties. Table 2 lists the parameters (gas density distribution,
rotation curve and local turbulent velocity dispersion) of our 
best disk model.
The position-velocity diagrams for the model disk and the observed CO
2$\rar$1 emission along P.A. = 40$\degrn$ (the dynamical major axis of the
central disk) are shown in Figure 4. The observed and model
line profiles are compared in Figure 5. The key ingredients of this model
are a bright central gas concentration at radii of up to $0\farcs5$,
a rising rotation curve to a maximum of $280 \pm 40\ \kms$ at R$\sim 0\farcs75$
(for an assumed inclination of 70$^{\circ }\ $--75$^{\circ }\ $) and a large 
turbulent velocity width of FWHM 300$\pm $ 50 $\kms$. The inferred 
model rotation curve as a function of radius is
plotted in Figure 7. The average ratio of rotation velocity in the
central disk to the one dimensional velocity dispersion ($\sigma \sim 0.43$
FWHM) is $<v_{rot}>/\sigma \sim 2$, with a range of 1 to 3. This indicates
that the central molecular disk is highly turbulent.  This, in turn,
implies that the disk is geometrically
thick.  These dynamics and the mass density (see below) are similar to those 
of stars in cores of elliptical
galaxies (Bender et al.~1994). In fact the model requires a
FWHM z-thickness of about $0\farcs7$ (330 pc) in order to 
match the profiles in the SE and NW corners of the central concentration
(Figure 5).  However, even with this large scale height the model profiles 
fit the data poorly S, SW and SE of the peak.  This probably indicates that 
separate kinematical systems, perhaps related to the streamers A through C 
that appear to merge here into the central disk, are present.  It is
impossible for any axisymmetric kinematic model to reproduce all of the
complicated kinematic and spatial structures observed in a violent
merger event such as that of NGC 6240.

The large local line widths and relatively small, large scale velocity
gradients suggest that radiative transport effects through the disk may play
a role. Following Downes and Solomon (1998) we have computed 
models based on the disk code of Dutrey, Guilloteau
and Simon (1994) with parameters appropriate for NGC 6240.  The 
code, which was modified by Downes and Solomon, includes non-LTE
radiative transport effects to account for finite 
opacity. For the relatively small optical depths that best fit the CO emission
characteristics of NGC 6240 (section 4 see below) the kinematic
parameters derived from the radiative transfer model are very similar to 
those listed in Table 2.

Since the gas dynamics are midway between a rotationally supported, cold
rotating disk and a dispersionally supported, hot system it is appropriate
to analyze the gas dynamics in terms of the observed velocity dispersion as
well. Figure 4 shows the second moment map, $\sigma _{v}$, 
\footnote{$\sigma _{v}=[\int (v^{2}-\langle v\rangle ^{2})I_{CO}(v)dv]/
[\int I_{CO}(v)dv]$} of the CO 2$\rightarrow $1 emission. Taking $\sqrt{2}
\sigma _{v}$ as a measure of the equivalent circular velocity, the
rotation curve inferred from this map is in excellent agreement with the
results of the rotating disk model in Table 2 (see Figure 7).
The dispersion increases from the CO flux peak to a maximum of 210
$\kms$ (equivalent circular velocity of 300 $\kms$) in an oval ring at 
projected radius $0\farcs5 - 1^{\prime \prime }$ and then
decreases slowly at larger radii. {\it We thus conclude from modeling 
the gas dynamics that there is a prominent concentration of mass centered 
on the CO flux peak (between the radio/infrared nuclei) which is extended 
over about $ 1^{\prime \prime }$}. In a separate paper (Tecza et al. 1999) 
we show from
sub-arcsecond resolution near-infrared field imaging spectroscopy that the
stellar dynamics in the center of NGC~6240 also supports this conclusion.
Tecza et al.~show that the stellar velocity dispersion peaks between the
two nuclei, even after they account for the rotation of the stars in the
individual nuclei.

While the axisymmetric disk model provides a satisfactory overall fit to 
the large scale structures observed in the p-v diagram (Figure 4), it
deviates significantly from the data in the central, low velocity emission
(Figure 5).
Compared to the model, the observed line profile is asymmetric and the
peak emission is blue-shifted by a few tens of $\kms$ with respect to the
velocity centroid of the outer, high velocity emission (Figure 4). This
discrepancy is mirrored in the channel maps, which show that the peak
emission is centered $\sim 0\farcs15$ north) of the line
connecting the peaks in the blue and red wings. The central disk is therefore
not strictly axisymmetric. There may also be additional spatial and 
kinematic substructure
of the gas very near the center and at the individual IR/radio nuclei which 
will require higher resolution.

\subsection{1.3 mm continuum map}

The spatial distributions of 1.315 mm (228 GHz) continuum and CO 2$
\rightarrow $1 line emission are clearly different (Figure 3). The continuum
emission is dominated by the slightly resolved southern nucleus (4.5 mJy, 19$
\sigma $ detection), with a second weaker component (1.4 mJy, 6$\sigma $
detection) on the northern nucleus. There is no obvious continuum peak on
the central CO concentration other than perhaps a shoulder NNE of the south
nucleus. If the non-thermal 8 GHz flux densities of the two nuclei (31 and
14 mJy, Colbert et al. 1994) are extrapolated with the radio spectral index
(between $-$0.6 and $-$0.7), one would expect 2.9--4.2 mJy and 1.3--1.9 mJy
at 1.315 mm for the southern and northern nuclei, respectively. These fluxes
densities are consistent with our measurements. The 1.3 mm emission from the
two nuclei, thus, is largely non-thermal synchrotron emission. Any continuum
source associated with the central CO concentration would have to be weaker
than 3 mJy within R$\leq 1^{\prime \prime }$. We discuss 
the lack of $\lambda = 1.3$ mm continuum emission from the central CO
concentration below.

\section{Gas Dominates the Central Mass Concentration}

Our CO J=2$\rightarrow $1 maps show a prominent concentration of molecular
gas centered in between the two radio/near-infrared nuclei. The gas
concentration is also the center of rotation and marks the maximum of
velocity dispersion in both the stars and the gas. The observational
evidence points to the CO peak coinciding with a mass
concentration between the nuclei.  In this section we show quantitatively
that this mass concentration is largely due to the (self-gravitating)
interstellar gas itself.

\subsection{Comparison of sizes of CO emission and mass concentration}

If the CO flux accurately traces the  distribution of molecular gas mass, 
one would qualitatively
expect that the radial distributions of the circular velocity\footnote{
the circular velocity v$_{c}(R)$ and the mass enclosed within R, M(R), are
related through $GM(R)=v_{c}^{2}(R)R$. This relationship holds exactly for a
spherical mass distribution, and approximately also for a disk-like
axisymetric distribution (Binney and Tremaine 1987).} and of the square root
of the ratio of CO flux within radius R divided by that radius are similar.
That is, the kinetic energy (rotational plus random) should vary with
radius as the ratio of the interior CO flux to the radius.
This is in fact the case (Figure 7, left).

\subsection{Gas mass estimates}

There are several approaches to estimating the gas mass quantitatively. A lower
limit to the gas mass can be obtained by assuming that the CO J=2$\rightarrow
$1 emission is optically thin. From excitation calculations with gas
temperatures between 50--100 K, local molecular hydrogen volume densities
between 10$^3$--10$^4$ cm$^{-3}$, a CO to H$_2$ fractional abundance ratio
of $8 \times 10^{-5}$ and a 38\% by mass contribution to the gas mass by
helium we find for the optically thin case 
\begin{equation}
M_{gas}\mbox{(CO thin)} = 2\times 10^6 L_{CO 2\rightarrow 1}
\mbox{ 
[Jy $\kms$]} D_{97 Mpc}^2 \quad\mbox{[M$_{\odot}$]}
\end{equation}
approximately independent of temperature. Taking the observed flux in the
central R$\le$1$^{\prime\prime}$ (528 Jy $\kms$), the equivalent optically
thin gas mass is $1.1\times 10^9$ M$_{\odot}$. However, neither the observed
peak brightness temperature (21 K), nor observed $^{12}$CO and $^{13}$CO
line ratios can be matched in this optically thin limit. Casoli et al. 
(1992) find that the 
$^{12}$CO $2\rightarrow 1/1\rightarrow 0$ brightness temperature line ratio
(corrected for beam and source size effects) is $\sim$0.8 and the $\frac{
^{12}CO 1\rightarrow 0} {^{13}CO 1\rightarrow 0}$ line ratio is $\sim 44 \pm
17$. Using a large velocity gradient excitation model we find that a match
to the $^{12}$CO 2$\rightarrow$1 brightness temperature, the $\frac{^{12}CO
2\rightarrow 1}{^{12}CO 1\rightarrow 0}$ line ratio, and the $\frac{^{12}CO
1\rightarrow 0}{^{13}CO 1\rightarrow 0}$ line ratio can be obtained with
molecular hydrogen column densities from 1 --4$\times 10^{19}$
cm$^{-2}$/($\kms$), which yields a $^{12}$CO 2$\rightarrow$1 line with 
$\tau \sim 1-4$. For this case the gas mass is 
\begin{equation}
M_{gas}\mbox{(radiative transport)} = 4.2\times10^6 L_{CO 2\rightarrow 1} 
\mbox{(Jy $\kms$)} D_{97 Mpc}^2 \quad\mbox{[M$_{\odot}$]}
\end{equation}
resulting in a mass of $2.2\times 10^9$ M$_{\odot}$ within the central 
R=1$\asec$. For self-gravitating, virialized molecular clouds at kinetic
temperature T and average molecular hydrogen density $\langle n(H_2) \rangle$
the gas mass (again 38\% helium) is also proportional to the integrated CO 2$
\rightarrow$1 flux in the optically thick limit (due to the mass dependence
of the overall line width; see Dickman et al. 1986, Genzel 1992) via 
\begin{equation}
M_{gas}\mbox{(GMCs)} = 3.1\times 10^7 L_{CO 2\rightarrow 1} \mbox{(Jy
$\kms$)} D_{97 Mpc}^2 (X/X_G) \quad\mbox{[M$_{\odot}$]}
\end{equation}
Here the conversion factor X between H$_2$ column density and CO intensity (I
$_{CO}$(K $\kms$) is given in units of the Galactic value appropriate for T$
\sim$10 K ,$\langle n(H_2) \rangle$$\sim$200 cm$^{-3}$, and X$_G = 2.3\times
10^{20}$ [cm$^{-2}$/(K $\kms$)] (e.g. Bloemen et al. 1989). Assuming 
that the
gas kinetic temperature in NGC~6240 is close to the average dust
temperature, T$\approx$T$_{dust} \sim$52 K (Solomon et al. 1997, Draine
1990) and that the average H$_2$ volume density is $\sim$200 cm$^{-3}$ we
find 
\begin{equation}
(X/X_G)_{NGC6240} = \left(\frac{10K}{T}\right) \left(\frac{\langle n(H_2)
\rangle} {200 \mbox{cm}^{-3}}\right)^{0.5} \sim 0.2
\end{equation}
We estimate mean H$_2$ densities of a few hundred H$_2$ cm$^{-3}$ from the
CO source size and the total gas mass (Table 1). Based on the argument that
in very luminous infrared galaxies, such as NGC 6240, the molecular gas is
not clumped into self-gravitating clouds but is distributed in an intercloud
medium with total gas masses less than the dynamical mass (Downes et
al.~1993), Solomon et al. (1997) and Downes and Solomon (1998) also infer 
X/X$_G$ $\sim$0.2 but find variations of factors of 3--4 from galaxy to galaxy.
With these assumptions equations 3 and 4 yield $M_{gas}$(GMCs)$\sim
3\times 10^9$ M$_{\odot}$ for the central R$\le$1$^{\prime\prime}$.

Draine (1990) has estimated the total dust mass and gas mass (gas-to-dust
mass ratio of 100) in NGC~6240 from dust emission models for the overall far-
infrared/submillimeter spectral energy distribution. Depending on the number
and temperature of dust components contributing to the emission he finds
total gas masses ranging between $6-70\times 10^{9}$ M$_{\odot}$. Assuming
that the spatial distribution of the far-infrared emission is similar to
that of the CO 2$\rightarrow$1 emission, about half of that mass, or $
3-35\times 10^9$ M$_{\odot}$, is expected to be in the central R$\le$1$
^{\prime\prime}$ region.

Finally, the optically thin 1.3 mm continuum emission can be used to derive
an upper limit to the gas mass (for M$_{gas}$/M$_{dust}$ = 100) 
\begin{equation}
M_{gas}\mbox{(1.3 mm)} = 3.8\times 10^8 S_{228}\mbox{[mJy]} D_{97 Mpc}^2
\nu^{-4} (0.0014/\kappa_{228}) (52 K/T_{dust})\quad\mbox{[M$_{\odot}$]}
\end{equation}
Here S$_{228}$ is the continuum flux density at frequency $\nu = 228$~GHz,
and $\kappa_{228}$ is the dust absorption coefficient at 228~GHz. Since the
1.3 mm continuum emission is almost certainly optically thin it has the
potential for being the most unambiguous method of determining the gas mass
in the central disk. However both $\kappa_{228}$ and the gas-to-dust mass
ratio are very poorly known. This severely increases the 0 of
masses derived in this manner. The value for $\kappa_{228}$ = 0.0014 cm$^2$
g$^{-1}$ used in equation 5 is taken from Draine and Lee (1984) and $\kappa$
scales with frequency as $\nu^2$. Taking the upper limit for thermal dust
emission associated with the central CO concentration estimated in section 3
($\le$3 mJy), equation (5) gives an upper limit to the gas mass of $
1.1\times 10^9$ M$_{\odot}$, which is close to the gas mass obtained for the
optically thin limit estimate, but at least two times smaller than the other
estimates for the CO emission. Table 3 summarizes the various mass
determinations.

All of the above mass estimates are intrinsically uncertain by factors of 2
to 3 due to the poorly known millimeter dust emissivities and CO flux to H$_2
$ column density conversion factors. With the exception of the even more
uncertain far-infrared estimates the agreement between the different mass
values in Table 3 is thus quite satisfactory. Leaving out the more uncertain
mass determined from fitting the far-infrared spectral energy distribution
(Draine 1990) the mean of all estimates gives a gas mass of $\approx 2\times
10^9$ M$_{\odot}$ in the central R=1$^{\prime\prime}$. This value is very
close to the radiative transport estimate (equation 2) suggesting that the
optical depth of the 1$\rightarrow$0 and 2$\rightarrow$1 $^{12}$CO lines is
moderate ($\tau$ of 1 to a few), and significantly smaller than in the
average molecular cloud medium in the disk of our own Galaxy, where the
typical CO optical depths range from 10 to 30. The difference plausibly lies
in the very large local line widths of the highly turbulent gas flows
throughout the central kiloparsec of NGC~6240, which is somewhat reminiscent 
of the broad line molecular cloud features in the center of our Galaxy.

The discrepancy between the best mass estimates from the CO emission and
that from the 1.3 mm continuum emission from dust is interesting, especially
since the latter is only an upper limit and there is no obvious continuum
peak at the CO concentration. It is possible that the dust temperature
in the high density core of the gas disk, being relatively far away from the
luminous nuclei, is significantly lower than the temperature characterizing
the far-infrared emitting dust grains. In fact the Draine (1990) modeling of
the overall infrared to millimeter spectral energy distribution of NGC~6240
suggests that the dust mass may be dominated by cooler, 20--25 K dust. In a
situation where cloud-cloud shocks play a very important role, gas and dust
temperatures may very well be decoupled. Dust temperatures from the cooler
cloud cores may then be significantly lower than the gas temperatures from
the shock heated cloud surfaces. In that case the limit inferred from
equation (5) would be 2 to 2.6 times greater, in much better agreement with
the best estimates from the CO emission.

\subsection{Mass Modeling}

Another means for determining the mass distribution in NGC~6240 can be
applied because of the good qualitative agreement between the spatial
distributions of dynamical mass and CO flux. Following the method of
Scoville et al. (1997) and Tacconi et al. (1999) we treat $\beta =X/X_{G}$
in equation (3) as a free parameter and solve for it by finding the best fit
of the sum of gas and stellar mass to the spatial distribution of the
dynamical mass. This approach exploits the fact that the dynamical mass
distribution has a characteristic bump at R=$0\farcs5 - 1^{\prime \prime }$ 
that needs to be matched. It is fairly straightforward to
estimate the mass of stars in the young, but aging, starburst of NGC~6240,
since this component is well traced by the K-band light. The deep stellar
absorption features (CO overtone bands, Mg, Ca, Na etc.) in the K-band
require an underlying early M- supergiant stellar population, and imply an
age of $1-3\times 10^{7}$ years for a short duration burst (Lester et al.
1988, Tezca et al. 1999). The low Br$\gamma $ equivalent width and
[NeIII]/[NeII] ratio are consistent with this estimate, and also require
that the starburst duration be significantly shorter than its age (Tezca et
al. 1999). With these constraints and a Scalo (1986) initial mass function
between 0.2 and 100 M$_{\odot }$, the mass to K-band luminosity ratio is 
$ \geq $0.5 (e.g. Thatte et al. 1997). With an effective V-band screen
extinction of 5 to 7 mag (or alternatively 40 mag of `mixed' extinction, see
Genzel et al.~1998) the mass of young stars in the starburst component is 
\begin{equation}
M_{\ast }\mbox{(young)}=9.3\times 10^{7}\left[\frac{M\ast /L(K)}{0.5}\right]
\left[\frac{\exp (0.09A(V))}{1.9}\right]S(K)\quad \mbox{[M$_{\odot}$]}
\end{equation}
where S(K) is the observed K-band flux density in mJy. Within R$\leq
1^{\prime \prime }$ S(K)$\sim $13.5 mJy so that M$_{\ast }$(young)$\sim
1.3\times 10^{9}$ M$_{\odot }$ (Tecza et al. 1999).

While this estimate shows that the southern and northern nuclei must be
fairly massive entities just based on their young star content, their masses
could be larger still if they are the nuclei of the progenitor galaxies with
a large additional old stellar component. The mass of the old stellar
component cannot be determined from the near- infrared data, however. We can
derive an upper limit to the gas mass fraction by simply neglecting the old
stellar contribution to the mass. In this case a good fit to the dynamical
mass distribution is obtained for $\beta \sim 0.35$ and M$_{gas}$/M$_{dyn}$$
\sim $0.75--0.85. Another, perhaps more plausible, approach is to assume
that an old stellar component is present and that its radial distribution is
similar to that of the young stars. If the underlying old stellar component
were similar to that of our own Galaxy the old stellar mass within 500 pc
would be $2.5-4\times 10^{9}$ M$_{\odot }$, so that M$_{\ast }$(young+old)$
\sim $ 3--4M$_{\ast }$(young). With this range of old to young stellar mass
ratios the best match to the shape of the mass distribution is obtained from
the CO rotation curve and second moment map with $\beta \sim 0.18-0.3$ and
M$_{gas}$/M$_{dyn}$$\sim 0.4-0.7$. The case of $\beta $=0.25 is shown in
Figure 7. Larger stellar masses and smaller values of $\beta $ fail to
reproduce the relative amplitude of the bump in the dynamical mass model. In
contrast smaller stellar masses and larger values of $\beta $ fit better.
For $\beta $=0.25 the gas mass in the central R=1$^{\prime \prime }$ then is 
$4\times 10^{9}$ M$_{\odot }$ and M$_{gas}$/M$_{dyn}$$\sim $0.6. This mass
estimate is listed in the last row of Table 3. {\bf Our mass model suggests 
that the gas mass fraction is even higher (i.e. $\sim$1) within $0\farcs3$ 
of the CO peak.}

Solomon et al. (1997) have estimated a dynamical mass from the
linewidth of their single dish CO J=1$\rar$0 spectrum.  In their work
they assume that the CO brightness temperature is the same as the
far-infrared blackbody dust temperature, compare that with the single
dish CO luminosity and derive a minimum CO emission radius.  Their
dynamical mass estimate of 10$^{10}$ M$_{\odot}$ in a 1$\asec$ radius is
nearly a factor of 2 larger than our derived estimate in Table 3.  The
discrepancy is due to the fact that the Solomon et al. estimate was
derived assuming that the entire single dish CO linewidth was due to the 
circular rotation of the gas within this radius.  Given the different
assumptions used by Solomon et al. and
this work to calculate the dynamical mass within the 
central R=1$\asec$ a factor of 2 agreement seems entirely reasonable.

\section{Fate of the Central Gas Concentration}

In the last section we have shown that a highly turbulent disk of
interstellar gas is in the process of settling within the central few
hundred parsecs of the merger NGC~6240, and constitutes a considerable
fraction (between 30--70\%) of the dynamical mass there. This key result of
the observations is not expected on the basis of recent N-body simulations
of mergers with gas. Barnes and Hernquist (1996) and Mihos and Hernquist
(1996) find that the gas component rapidly dissipates angular momentum and
forms rotating central disk/bar concentrations on scales $\leq $1 kpc around
the progenitor nuclei. Such rapidly rotating, cold circum-nuclear disks
are, in fact, observed in Arp 220 and several other ULIRGs (Scoville et al.
1997, Downes and Solomon 1998, Sakamoto et al. 1999). However, the
simulations do not show cases where the gas becomes self-gravitating and
forms a central concentration between the nuclei, as we observe in NGC~6240.

The difference in the molecular spatial distributions and other
characteristics between NGC 6240 and several of the other ULIRGs (including
Arp~220) may be understandable in terms of NGC~6240 being in an earlier
stage of merger evolution than the other studied galaxies.  Simulations
by Barnes and Hernquist do show that, shortly after the first encounter
of two equal mass, gas-rich galaxies, a significant amount of gas can be
ram-pressure stripped from the disks of the progenitors and reside in the 
interface region between the disks.  During this phase no stellar mass 
is expected in this region, however, and this phase is relatively
short-lived as the galaxies fall back toward eachother for the next
encounter.  
The projected
linear separation of the two nuclei of NGC~6240 (730 pc) is a factor of
2 greater than that of Arp 220 (350 pc). In Arp 220 the true nuclear
separation may be $\sim $1.5 times the observed one (550 pc; Scoville et
al. 1997; Downes and
Solomon, 1998). In NGC~6240 the observed differential velocity 
between the two nuclei seen in both optical and near-infrared
spectroscopy is $\sim$100--150 km s$^{-1}$ (Fried and Schulz 1983; Tecza
et al. 1999).
This indicates that either the true nuclear separation could be
much larger than the projected one, between 1.5--3.5 kpc, 
or that the nuclei are merging on radial, rather than circular
orbits.  The former conclusion would further strengthen our finding that, in
the central R=1$^{\prime \prime }$ centered on the CO peak, gas, and not the
stars, dominates the mass. Much of the stellar mass included in our mass
modelling in section 4.3 may actually be far outside the central region. 

The conclusion that NGC~6240 could be in an earlier merging stage than Arp~220
and other ULIRGs is consistent with its having a lower luminosity. The
simulations of Mihos and Hernquist (1996) predict that in the merging of
disk galaxies with bulges there are several starburst episodes associated
with each peri-passage of the two galaxies. The most luminous starburst
occurs in the last phase when the two nuclei merge into a single one,
and NGC~6240 is likely to be in this final phase given the fact that the
two close nuclei will probably not separate again.

\subsection{Rapid Dissipation of Random Motions}

NGC~6240 stands out among luminous infrared galaxies by having
the most luminous infrared line emission from hot molecular hydrogen. For a
wide range of physical conditions the brightest H$_2$ infrared
line is the v=0-0 S(1) line at 17$\mu$m. The dereddened luminosity of that
line in NGC~6240 is $3\times 10^8$ L$_{\odot}$ (Egami et al. 1999). ISO and
ground-based observations of a number of rotational and ro-vibrational H$_2$
lines show that the total luminosity in all infrared H$_2$ lines is at least
7 times greater than that of the 17$\mu$m S(1) line (Tecza et al. 1999;
Egami et al. 1999), so L(H$_2$)$\sim 2\times 10^9$ L$_{\odot}$. This
corresponds to 0.3\% of the bolometric luminosity of NGC~6240. By
comparison, the fractional H$_2$ luminosity in Arp~220 and a sample of about
a dozen other luminous infrared galaxies is between 2 and 10 times smaller
(Rigopoulou et al. 1999). Another galaxy with comparable L(H$_2$)/L(FIR)
is the Antennae (NGC~4038/39) system. In this colliding galaxy system the
nuclei have a separation of $\sim$6 kpc, and there is also a prominent
concentration of molecular gas in the interaction region between the nuclei
(Stanford et al. 1990).

Based on the large 2$\mu$m 1$\rightarrow$0 S(1) H$_2$/Br$\gamma$ line ratio,
van der Werf et al. (1993) have concluded that the most likely process
exciting the H$_2$ line emission in NGC~6240 is slow ($\le 40 \kms$) C-shocks
in cloud-cloud collisions. This is supported by the H$_2$ level populations
derived from the recent ISO observations of Egami et al.~(1999). In such
shocks line emission from a number of species other than H$_2$ (OH, CO,
[OI]) is expected to be significant so that L(H$_2$) is a lower limit to the
total cooling rate.

From the intense H$_{2}$ line emission it follows that the large local
random velocities must dissipate rapidly. The dissipation time $\tau _{diss}$
within R$\leq $470 pc is then given by 
\begin{equation}
\tau _{diss}\le 1.5M_{gas}\sigma ^{2}/L(H_{2})\sim 10^{6.8}M_{9.3}\sigma
_{130}^{2}/L(H_{2})_{9.3}\quad \mbox{[yrs]}
\end{equation}
where $\sigma _{130}$ is the average local velocity dispersion in units of
130 $\kms$, M$_{9.5}$ is the gas mass in units of $3\times 10^{9}$ M$_{\odot }$
, and L(H$_{2}$)$_{9.3}$ is the H$_{2}$ luminosity in units of $2\times
10^{9}$ L$_{\odot }$. The dissipation time scale in the central disk thus is
only about 2 to 3 dynamical time scales! We conclude that the present, highly
turbulent flow will probably settle to a dynamically cold disk supported
largely by rotation in $<10^{7}$ years. This time is significantly shorter
than the present orbital time scale of the two nuclei ($\geq 10^{7.5}$
years) that determines the `clock' of the merger.

\subsection{Onset of star formation in the central gas disk}

One of the surprising results of this study of NGC~6240 is that, despite the
very high central H$_2$ column densities (N(H$_2$)$\sim 1-2\times 10^{23}$ cm
$^{-2}$, $\Sigma \sim 10^{3.6}$ M$_{\odot}$ pc$^{-3}$) most of the present star
forming activity is occurring in the two nuclei as shown by the radio
continuum map of Colbert et al. (1994) and the K-band and Br$\gamma$ maps of
Tecza et al. (1999). However, the basic `Toomre criterion' (1964) for a
rotating disk with rotation velocity v$_{rot}$, and effective `sound speed' 
v$_s$ requires that for a disk of surface density $\Sigma$ to be
gravitationally unstable, v$_s$ must be smaller than 
\begin{equation}
v_s \le \zeta (G \Sigma R/v_{rot}) \sim 11 \Sigma_{23}
R_{0\farcs5} /v_{250} \quad\mbox{[$\kms$]}
\end{equation}
where $\Sigma_{23}$ is the H$_2$ column density in units of 10$^{23}$ cm$
^{-2}$, v$_{250}$ is the rotation velocity in units of 250 $\kms$ and $\zeta$
is a dimensionless number between 1.1 and 1.6 (see chapter 5.3 in Binney and
Tremaine 1987). For $v_s \sim \sigma \sim$130 $\kms$ the disk cannot locally
collapse and form stars and the effective Jeans mass is very large (a few$
\times 10^5$ M$_{\odot}$). Only when the dispersion has dissipated to less
than 10 $\kms$ can local gravitational instability set in.

We thus conclude that in a few dynamical time scales ($\le 10^7$ years), and
perhaps triggered by the next peri-passage of the two nuclei, NGC~6240 may
experience a major starburst in the central CO disk and more closely
resemble Arp~220 and other ULIRGs. NGC~6240 and Arp~220, thus, support the
scenario (Mihos and Hernquist 1996, Genzel et al. 1998, Genzel, Lutz and
Tacconi 1998) that mergers go through several short phases of very active
star formation. It remains unclear for NGC~6240 during this process what
fraction of the gas mass then will be in the general central gas disk and
how much will collect in smaller circumnuclear disks around the progenitor
nuclei. In Arp~220 this ratio is about 1:1 (Downes and Solomon 1998;
Sakamoto et al. 1999).

\section{Conclusions}

From our high resolution study of the molecular gas in NGC~6240 we have
found:

\begin{itemize}
\item  {About half of the CO flux is concentrated in a thick disk-like
structure located between the IR/radio nuclei. The velocity structure of
this disk is highly disturbed, with unusually large line widths of FWZP
700--1000 $\kms$ almost everywhere within the central 3$^{\prime\prime}$ from
the CO peak.}

\item  {Modeling of the data shows that the gas is rotationally supported,
but with a very large local velocity dispersion characteristic of a hot
dynamical system. The inferred average ratio of rotational velocity in the
central disk to local one dimensional velocity dispersion is $\langle
v_{rot}\rangle /\sigma \sim 2$ with a range of 1 to 3. This indicates that
the disk is highly turbulent and geometrically thick.}

\item  {Based on several methods to estimate the gas mass and mass models
for NGC~6240 we find that the mass of the gas disk within the central R$\leq 
$1$^{\prime \prime }$ is $2-4\times 10^{9}$ M$_{\odot }$, and constitutes
between 30\% and 70\% of the dynamical mass in this region. Self-gravitating
gas concentrations thus may play a significant role in the evolution of
mergers.}

\item  {NGC~6240 is likely in an earlier merger stage than typical ULIRGs.
From the very high luminosity in the infrared H$_{2}$ lines we conclude that
the local random velocities will dissipate in a few dynamical times 
($<10^7$ years). Once this happens, NGC~6240 will form a dense,
geometrically thin central gas disk and will be ripe for a major starburst.
With its luminosity then rising and its ISM more confined, it could then 
more closely resemble Arp~220 and other ULIRGs.}
\end{itemize}

Acknowledgements. We thank the staff of IRAM for their help with carrying
out the observations and calibrations. We are grateful to E.Colbert for
providing us with a FITS image of his 8 GHz radio map.  We also thank R.Bender
for interesting discussions, and the referee, Chris Mihos, for many 
useful suggestions which helped to improve the final version of this
paper.

\clearpage

\figcaption{Selected CO J=2$\rar$1 line spectra (20 $\kms$ resolution) 
superposed on a false color image of the integrated CO 2$\rar$1 flux 
distribution (logarithmic stretch). The slightly overresolved synthesized
beam with $0\farcs5\times 0\farcs7$ FWHM resolution is shown as a red
ellipse in the lower left. Letters A through E mark the gas features 
discussed in section 3 of the text. The two asterisks mark the positions 
of the south and north radio/infrared nuclei, and the crosses mark the 
positions of the spectra. \label{fig1}}

\figcaption{Uniformly weighted velocity channel maps of the CO J=2$\rar$1 
emission in steps of 40 $\kms$ and with a spatial resolution of $0\farcs5
\times 0\farcs9$ FWHM.  The cross denotes the position of the CO flux
peak.  Contours run from $-$18 to $-6$ mJy beam$^{-1}$ and  
from 6 to 36 mJy beam$^{-1}$ in steps of 6 mJy beam$^{-1}$ ($\sim 3\sigma$ 
per 40 $\kms$ channel). \label{fig2}}

\figcaption{Left: 1.315mm (228 GHz) continuum map at $0\farcs5 \times 0\farcs9$
FWHM resolution. 
Contours start at 0.7 mJy beam$^{-1}$ ($\sim 3\sigma$) and increase in steps of 
0.7 mJy beam$^{-1}$. The 
cross denotes the position of the CO flux peak. Right: Redshifted CO emission 
(intensity weighted sum from 345 to 545 $\kms$, thin contours) and blueshifted 
CO emission (intensity weighted sum from $-$535 to $-$335 $\kms$, heavy 
contours).  Contours run in steps of 10\% of the peak in each case.  The
cross denotes the position of the CO flux peak. \label{fig3}}

\figcaption{Top: Contour and grey scale map of the CO 2$\rar$1 velocity 
dispersion. 
Contours run from $\sigma_{moment}$=60 to 200 $\kms$ in steps of 20 $\kms$. 
Middle: Position-
velocity map of the CO 2$\rar$1 emission along the direction of largest 
velocity gradient (P.A.=40$\degr$) in the central disk. Bottom: 
Position -velocity map of the best 
fitting axisymmetric rotating disk model (thin contours, Table 2) superposed 
on the data (heavy contours). \label{fig4}}

\figcaption{Overlay of model profiles of rotating disk (thin line; parameters 
as in Table 2) on CO 2$\rar$1 spectra (thick lines), for selected positions 
in the central CO disk.
Data and model are at a resolution of $0\farcs5 \times 0\farcs9$
resolution.  Velocities are 
relative to 7389 $\kms$. Position offsets from the CO peak are given in 
parentheses in the top left of each box. \label{fig5}}

\figcaption{Top: Overlay of a logarithmically spaced contour map of the 8~GHz 
radio continuum emission (resolution $0\farcs15$ FWHM, Colbert et al. 1994)
on a 
false color map of the CO 2$\rar$1 integrated flux (logarithmic stretch). 
Radio contours run from $10^{-1}$ to $10^{1.2}$ mJy in steps of 10$^{0.2}$. 
Middle: Overlay of a logarithmically spaced contour map of the 5~GHz
MERLIN radio continuum emission (resolution $94\times 50$~milliarcsec at
P.A. 21$\degrn$) on the false color 
linear scale CO map to emphasize the central gas disk and the
compactness of the radio nuclei. Bottom: 
Logarithmic false color 8 GHz radio continuum map (as before, Colbert et 
al. 1994) with redshifted (intensity weighted sum from 345 to 545 $\kms$, 
thin red contours) and blueshifted (intensity weighted sum from $-$535 to 
$-$335 $\kms$, heavy light blue contours) CO 2$\rar$1 emission contours 
superposed. The cross denotes the position of the CO flux 2$\rar$1 peak. 
\label{fig6}}

\figcaption{Left: CO model rotation curve (Table 2) shown as a heavy curve; 
CO 2$\rar$1 second moment 
map (CO dispersion $\sigma_{moment}$) converted to circular velocity,
(v$_{circ}$(CO) $\sim\protect{\sqrt{2}}\sigma_{moment}$, see footnote 5), 
plotted as
crosses and connected by thin lines; and the square root of the observed 
CO flux distribution divided by radius in dashed lines scaled to match. Right: 
Mass models in the central $1\farcs4$ (see text).  The light
solid curve shows the stellar mass for the case M$_{old}$ + M$_{young}$ 
$\approx 3\times$M$_{young}$; the medium solid curve is the distribution of
M$_{gas}$ assuming the CO-to-H$_2$ conversion factor is 0.2 times the Galactic
value of Bloemen et al. (1989); the dashed line shows the model dynamical
mass distribution; the filled squares represent dynamical mass distribution
derived from the CO moment map; and the heavy solid curve shows the total of
the observed gas and stellar mass. For this case
M$_{gas}$/M$_{dyn}$$\approx$0.5.  Better fits to the characteristic
shape of the dynamical mass distribution can be achieved with larger
M$_{gas}$/M$_{dyn}$ ratios and correspondingly smaller contributions of
old stars in the central R$\le$1$\asec$.\label{fig7}}

\clearpage

\begin{deluxetable}{cc}
\tablenum{1}
\tablewidth{0pt}
\tablecaption{Basic Observational Parameters}
\tablehead{}
\startdata
Total CO 2$\rar$1 flux& 1220 Jy $\kms$ \nl
CO 2$\rar$1 flux in central R=1$\asec$& 528 Jy $\kms$ \nl
Peak CO 2-1 T$_b$& 21 K \nl
228 GHz cont.~flux - south nucleus&4.7 mJy \nl
228 GHz cont.~flux - north nucleus& 1 mJy\nl
\enddata
\end{deluxetable}

\clearpage
\begin{deluxetable}{cc}
\tablenum{2}
\tablewidth{0pt}
\tablecaption{Best Fitting Parameters of Axisymmetric Disk Model}
\tablehead{}
\startdata
R (arcsec)& v$_{rot}$ ($\kms$)\nl
0.2& 120\nl
0.3& 180\nl
0.5& 280($\pm$40)\nl
0.75& 280\nl
1.0& 250\nl
Density distribution& n(R)=$\exp{[-2.77((R-0.1)/0.55)^2]}$\nl
\ &$ + 0.25\times\exp{(-2.77(R/1.5)^2)}$\nl
\ &(between 0.01$\asec$ and 2$\asec$)\nl 
Local velocity dispersion& $\sigma$=128 $\kms$\nl
FWHM z-thickness of disk& $0\farcs7$
\enddata
\end{deluxetable}

\clearpage
\begin{deluxetable}{lc}
\tablenum{3}
\tablewidth{0pt}
\tablecaption{Gas Mass Estimates for the Central CO Concentration}
\tablehead{
\colhead{Mass Estimate} & \colhead{Gas Mass Within R=1$\asec$ (470 pc)}}
\startdata
Optically thin CO emission& $1.1\times 10^9$ M$_{\odot}$\nl
Radiative transport solution matching&$2.2\times 10^9$ M$_{\odot}$\nl
$^{12}$CO 2$\rar$1, 1$\rar$0 and $^{13}$CO 1$\rar$0 lines&\ \nl
Optically thick CO emission&$3\times 10^9$ M$_{\odot}$\nl
Far-infrared dust emission& $10(+25,-7)\times 10^9$ M$_{\odot}$\nl
1.3mm dust emission& $\le 1.1\times 10^9 (52/T_{dust})$ M$_{\odot}$\nl
Mass modeling of CO emission&$4\pm 1 \times 10^9$ M$_{\odot}$\nl 
Dynamical mass& $6.4 \pm 1\times 10^9$ M$_{\odot}$\nl
\enddata
\end{deluxetable}
\clearpage

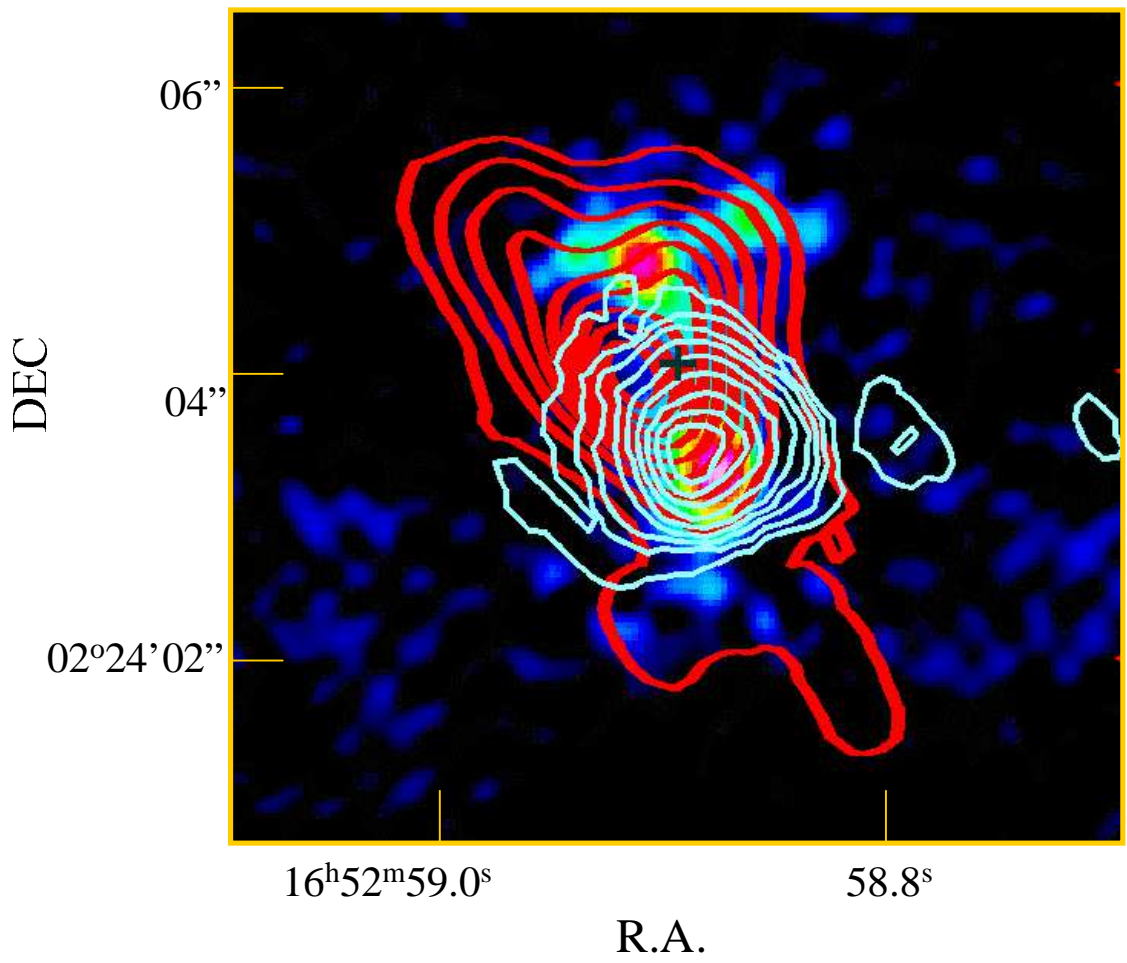
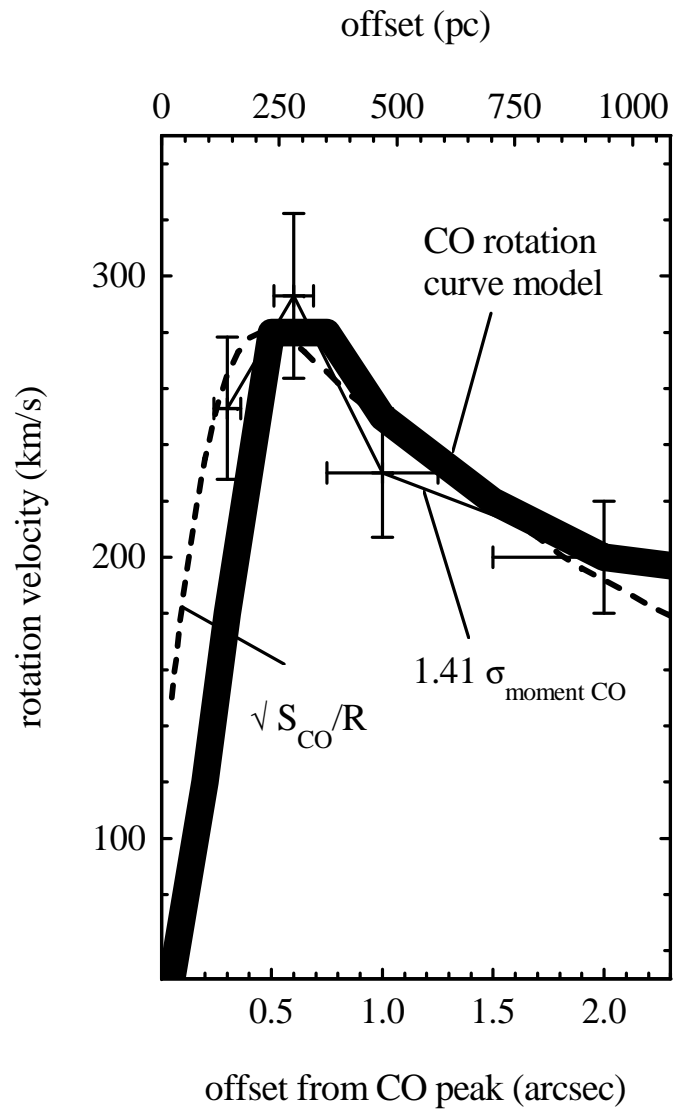
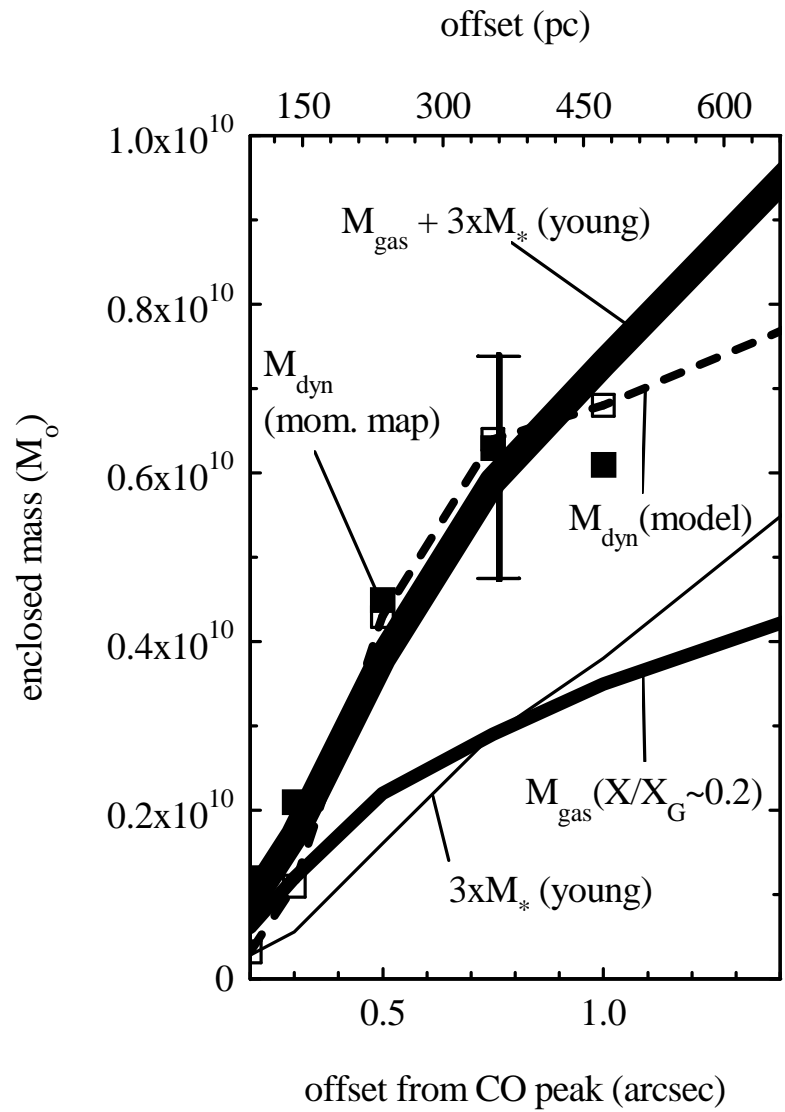
\end{document}